\documentclass[prb,twocolumn]{revtex4-1}

\usepackage{hyperref}
\usepackage{graphicx}
\usepackage{fullpage}
\usepackage{bbm,amssymb}

\def\ii{{\rm i}}
\def\ave#1{\langle #1 \rangle}
\def\tr{\,{\rm tr}\,}
\def\sx{\sigma^{\rm x}}
\def\sy{\sigma^{\rm y}}
\def\sz{\sigma^{\rm z}}

\def\tit#1{{\em #1},}
\def\etal#1{#1}

\begin{document}
\title{Transport in a disordered tight-binding chain with dephasing}
\author{Marko \v Znidari\v c and Martin Horvat}
\affiliation{Physics Department, Faculty of Mathematics and Physics, University of Ljubljana, Ljubljana, Slovenia}

\begin{abstract}
We study transport properties of a disordered tight-binding model (XX spin chain) in the presence of dephasing. Focusing on diffusive behavior in the thermodynamic limit at high energies, we analytically derive the dependence of conductivity on dephasing and disorder strengths. As a function of dephasing, conductivity exhibits a single maximum at the optimal dephasing strength. The scaling of the position of this maximum with disorder strength is different for small and large disorder. In addition, we study periodic disorder for which we find a resonance phenomenon, with conductivity having two maxima as a function of dephasing strength. If disorder is nonzero only at a random fraction of all sites, conductivity is approximately the same as in the case of a disorder on all sites but with a rescaled disorder strength.
\end{abstract}

\pacs{72.10-Bg, 03.65.Yz, 05.60.Gg, 75.10.Pq, 72.10-d}

\maketitle

\section{Introduction}

Understanding quantum transport in simple systems is of obvious relevance for understanding nano and mesoscopic devices as well as transport in real materials. One such model system is a tight-binding chain in the presence of disorder and dephasing, that has been studied already more than 30 years ago~\cite{Thouless:81}. The goal is to understand the interplay of disorder, incoherent processes and possibly interactions, on the transport properties. If the system is coherent procedure is, at least in principle, straightforward -- one has to calculate the transmission by one of various approaches, see e.g. book~\cite{Datta:95} for an overview. Non-coherent processes due to interaction with external degrees of freedom, for instance due to electron-phonon scattering, are more difficult to account for. Correspondingly, they are also less understood. In the present work we study the influence of the environment, in particular its dephasing effects, on the bulk quantum transport in a disordered system.

There are several approaches how to study incoherent transport. A rigorous one in terms of nonequilibrium Green's function is difficult to analytically evaluate for all but the simplest systems, while its numerical calculation is very time-consuming and limited to small systems. With that in mind, alternative, more phenomenological approaches to incoherent transport are actively investigated. One of the earliest suggestions accounts for the dephasing by introducing fictitious reservoirs that break phase coherence~\cite{Buttiker:85}. This so-called B\" uttiker probe has been applied to the disordered tight-binding model~\cite{Amato:90} and is successfully used for the calculation of transmission through molecules, see e.g. Refs.~\cite{Nozaki:08,Guo:09,Nozaki:12}. A phenomenological model of dephasing can also be constructed by using an appropriate self-energy in a nonequilibrium Green's function~\cite{Datta:07}. Another approach consists of treating the system as being composed of statistically distribu\-ted coherent and incoherent parts (having an imaginary self-energy)~\cite{Zilly:09,Zilly:12,Stegmann:12}. Dephasing can also be accounted for by simulating vibrational motion of individual sites~\cite{Kogan:10}, thereby modulating inter-site couplings, as is for instance done using molecular dynamics simulations in carbon nanotubes~\cite{Ishii:10}.

In this paper we use different approach by writing quantum master equation of the Lindblad type~\cite{lindblad} describing time evolution of the reduced density matrix of a tight-binding model without environment. Environmental dephasing is described in an effective way by dissipative operators that cause the decay of off-diagonal coherences (i.e., also of the current). Coupling the system in addition to reservoirs a true nonequilibrium steady state (NESS), reached after long time, is studied. Description with the Lindblad equation has its advantages and disadvantages. Compared to other approaches, once one finds the NESS one has access to all many-body observables, i.e. to a complete state. Of course, this advantage can be harnessed only if one is able to solve for the NESS. Fortunately, a tight-binding model with dephasing is solvable~\cite{JSTAT10} in its Lindblad formulation. While a study of NESSs of Lindblad equations is a well-defined and important mathematical problem it has its down sides also. A number of approximations, like weak coupling and fast-decaying bath correlations, is needed in standard derivations of the Lindblad equation~\cite{book}. We note, however, that Lindblad master equation is expected to provide good description on timescales that are much larger than the timescale of bath correlations. It namely provides the most general description for quantum processes that preserve positivity and trace and form a dynamical semigroup. Memory effects, not accounted for by the Lindblad equation, are expected to be smaller at higher temperatures, which is the regime that we study. In addition, we are interested in the bulk transport properties in the thermodynamic limit for which finite-size boundary effects are expected to play no role. Compared to small systems, where more complicated master equations might be necessary for the correct description of transport, see e.g.~\cite{Timm:08}, we can use simpler Lindblad equation to calculate bulk conductivity at high temperature.

Tight-binding model with dephasing and disorder has recently become of interest also in the context of excitation transfer in biomolecules, in particular in photosynthetic complexes. There a coupling with the environment, i.e., dephasing, increases the efficiency of excitation transfer because it counteracts localization due to disorder~\cite{Plenio:08,Rebentrost:09,Caruso:09,Hoyer:10,Kassal:12}. This, so-called environmentally assisted transport, can be in long homogeneous chains explained by analytical results that we present.

Our study proceeds in two steps. First, using the method introduced in Refs.~\cite{JSTAT10,PRE11} and extending it to an inhomogeneous system, we transform the problem from the state space of many-body density matrices, that is exponentially large in the chain length, to the one of two-point correlation functions, whose size is only quadratic in the chain length. We manage to analytically solve the resulting equations for some parameter regimes, while for other we resort to numerical solution. In addition to explaining the dependence of conductivity, i.e., of diffusion constant, on disorder and dephasing strength we also identify an interesting resonance phenomenon for periodically placed disorder.

\section{XX chain with dephasing}
We shall first present the model in the framework of Pauli matrices and show later a well-known fact that the model is equivalent to a nonequilibrium tight-binding model with an on-site disorder and experiencing dephasing due to environment.

The Hamiltonian of the XX spin chain in an inhomogeneous magnetic field is given by
\begin{equation}
{\cal H}=\sum_{j=1}^{L-1} (\sigma_j^{\rm x} \sigma_{j+1}^{\rm x} +\sigma_j^{\rm y} \sigma_{j+1}^{\rm y})+\sum_{i=1}^L \epsilon_i \sigma_i^{\rm z},
\label{eq:H}  
\end{equation}
with standard Pauli matrices and $L$ being the chain length. We use units in which $\hbar=1$ and nearest-neighbor hopping is $J=1$. We shall describe evolution of the system, described by a many-body density matrix $\rho$ of size $2^L$, in the presence of environment by Lindblad equation~\cite{lindblad},  
\begin{equation}
\frac{{\rm d}}{{\rm d}t}{\rho}=\ii [ \rho,{\cal H}]+ {\cal L}^{\rm dis}(\rho)\equiv{\cal L}(\rho),
\label{eq:Lin}
\end{equation}
in which dissipative influence of the environment is described by the so-called dissipator ${\cal L}^{\rm dis}$. Action of a dissipator on an arbitrary operator $\rho$ is expressed in terms of Lindblad operators $L_k$ as,
\begin{equation}
{\cal L}^{\rm dis}(\rho)=\sum_k \left( [ L_k \rho,L_k^\dagger ]+[ L_k,\rho L_k^{\dagger} ] \right).
\end{equation}
Dissipator will in our case consist of two parts, ${\cal L}^{\rm dis}={\cal L}^{\rm bath}+{\cal L}^{\rm deph}$. The first ${\cal L}^{\rm bath}$ describes the action of two baths, one at each chain end, and will induce a nonequilibrium situation with current flowing through the system. The bath is written as a sum of a part acting only at the left end (site index $j=1$ and label ``L'') and a part acting only at the right end (site index $j=L$ and label ``R''), ${\cal L}^{\rm bath}={\cal L}^{\rm bath}_{\rm L}+{\cal L}^{\rm bath}_{\rm R}$. Two Lindblad operators at the left end are
\begin{equation}
L^{\rm L}_1=\sqrt{\Gamma(1+\mu)}\,\sigma^+_1,\quad L^{\rm L}_2=\sqrt{\Gamma(1-\mu)}\, \sigma^-_1,
\label{eq:Lbath}
\end{equation}
while at the right end we have
\begin{equation}
L^{\rm R}_1=\sqrt{\Gamma(1-\mu)}\,\sigma^+_L,\quad L^{\rm R}_2=\sqrt{\Gamma(1+\mu)}\, \sigma^-_L,
\end{equation}
$\sigma^\pm_j=(\sigma^{\rm x}_j \pm {\rm i}\, \sigma^{\rm y}_j)/2$. Bath therefore flips the boundary spin up or down with certain probability. Parameter $\mu$ plays the role of the forcing, trying to induce magnetization $+\mu$ at the left end and $-\mu$ at the right end~\cite{foot1}. The other dissipative part, ${\cal L}^{\rm deph}$, represents the influence of environment at each site, being for instance due to scattering on phonons. The dephasing part ${\cal L}^{\rm deph}=\sum_{j=1}^L{{\cal L}^{\rm deph}_j}$ is a sum of ${\cal L}^{\rm deph}_j$, each of which acts only at the $j$-th site and is described by a single Lindblad operator, 
\begin{equation}
L^{\rm deph}_j=\sqrt{\frac{\gamma}{2}}\sigma^{\rm z}_j.
\label{eq:Ldeph}
\end{equation}
Due to dissipation the system's density matrix $\rho(t)$ will after long time converge to a time-independent state $\rho_\infty \equiv \lim_{t \to \infty} \rho(t)$ that is called a nonequilibrium steady state (NESS). For our model the NESS $\rho_\infty$ is unique and can be formally expressed via the solution of Lindblad equation (\ref{eq:Lin}) as $\rho_\infty = \lim_{t \to \infty} \exp{({\cal L}t)}\rho(0)$. In this work we shall be interested in the NESSs bulk transport properties in the presence of disorder and dephasing. Conductivity is expected to be insensitive to details of bath Lindblad operators in the thermodynamic limit. All expectation values reported in the paper are with respect to the NESS state, e.g. $\ave{A}=\tr{(A \rho_\infty)}$. 

Because we are interested in transport properties the central object we shall consider is the spin (i.e., magnetization) current $j_r$, defined at site $r$ through a continuity relation, resulting in expression $j_r \equiv \ii \, [\sz_r, \sigma_r^{\rm x} \sigma_{r+1}^{\rm x} +\sigma_r^{\rm y} \sigma_{r+1}^{\rm y} ]$, giving
\begin{equation}
j_r=2(\sx_r \sy_{r+1}-\sy_r \sx_{r+1}).
\end{equation}
Dephasing alone causes an exponential decay of off-diagonal matrix elements (in the standard basis in which $\sz$ is diagonal). Therefore, due to ${\cal L}^{\rm deph}(j_r)=-4\gamma j_r$, if there would be no ${\cal H}$ and no ${\cal L}^{\rm bath}$ in the Lindblad equation, dephasing would cause an exponential decay of current with time, $j_r(t)=\exp{(-4\gamma t)} j_r(0)$. In the presence of Hamiltonian and driving the evolution is more complicated, resulting in a nontrivial NESS. Evaluating the action (\ref{eq:Lin}) of dephasing (\ref{eq:Ldeph}) on the identity and $\sz$ operators, we see that ${\cal L}^{\rm deph}(\sz_r)={\cal L}^{\rm deph}(\mathbbm{1}_r)=0$ hold.

\section{Stationary solution}
To find the NESS $\rho_\infty$ we have to find stationary solution of Eq.~(\ref{eq:Lin}), i.e., solve ${\cal L}(\rho_\infty)=0$. Expanding $\rho_\infty$ in an operator basis the stationary equation can be written as a set of coupled linear equations for unknown expansion coefficients. Because the number of unknown coefficients is $4^L-1$, and therefore grows exponentially with the system size $L$, it is for a generic system impossible to analytically obtain $\rho_\infty$. XX model with dephasing though has a nice property that these exponentially many equations split into smaller sets of equations that are uncoupled~\cite{JSTAT10,PRE11,eisler11,temme:12}. Namely, there is a hierarchy of observables according to the number of fermionic operators they contain. For instance, all two-point observables decouple from the rest, meaning that one can write a closed set of equations for observables involving two fermionic operators. Once these are known, they can be used as a ``source'' term in equations for three-point observables, and so on for higher order correlations. For general consideration when such a hierarchy appears see Ref.~\cite{eisler11,Giedke:12}.

As shown in Ref.~\cite{JSTAT10}, one consequence of such structure is that the NESS can be written as
\begin{eqnarray}
\rho_\infty&=&\frac{1}{2^L}\left[ \mathbbm{1}+\mu\cdot(H+B) +{\cal O}(\mu^2)\right] \nonumber \\
H &=&\sum_{r=1}^{L}\sum_{j=1}^{L+1-r} h_j^{(r)} H^{(r)}_j \nonumber \\
B &=&\sum_{r=2}^{L}\sum_{j=1}^{L+1-r} b_j^{(r)} B^{(r)}_j,
\label{eq:ansatz}
\end{eqnarray}
where we define $H^{(r+1)}_j \equiv \sigma_j^{\rm x}Z_{j+1}^{(r-1)}\sigma_{j+r}^{\rm x}+\sigma_j^{\rm y}Z_{j+1}^{(r-1)} \sigma_{j+r}^{\rm y}$ and $B^{(r+1)}_j \equiv \sigma_j^{\rm x}Z_{j+1}^{(r-1)}\sigma_{j+r}^{\rm y}-\sigma_j^{\rm y}Z_{j+1}^{(r-1)} \sigma_{j+r}^{\rm x}$, both for $r\ge 2$ and with $Z_j^{(r)}\equiv \sz_j \cdots \sz_{j+r-1}$ being a string of $r$ consecutive $\sz$'s. $H_j^{(1)}$ is defined as $H_j^{(1)}\equiv - \sz_j$. Let us stress that Eq.~(\ref{eq:ansatz}) is not just a perturbative expansion in $\mu$ as it might seem at a first glance -- it is an exact ansatz~\cite{JSTAT10,PRE11} holding for any driving $\mu$. Or, in other words, terms proportional to $\mu^2$ (and of higher order) not written explicitly in Eq.~(\ref{eq:ansatz}) are all orthogonal (using Hilbert-Schmidt inner product $\langle A,B\rangle\equiv \tr{(A^\dagger B)}$) to operators in $H$ and $B$. Expectation value of these observables is therefore obtained exactly to all orders in $\mu$ by calculating only the expansion coefficients contained in $H$ and $B$. In particular, noting that $H^{(2)}_r$ is a hopping contribution to the energy density, $B^{(2)}_r$ current and $H^{(1)}_r$ magnetization, all quantities necessary to study transport are contained in the ansatz (\ref{eq:ansatz}).

Writing now ${\cal L}(\rho_\infty)=0$ using the ansatz (\ref{eq:ansatz}), we get the aforementioned closed set of linear equations for unknown coefficients $h_j^{(r)}$ and $b_j^{(r)}$. These are
\begin{eqnarray}
\Gamma+\Gamma h_1^{(1)}-b_1^{(2)}=0, \nonumber \\
-\Gamma +\Gamma h_L^{(1)}-b_{L-1}^{(2)}=0, \nonumber \\
b_j^{(2)}-b_{j+1}^{(2)}=0, \quad j=2,\ldots,L-2,
\label{eq:eqs0}
\end{eqnarray} 
and for $r \ge 2$
\begin{eqnarray}
\label{eq:eqs}
(h_j^{(r-1)}-h_{j+1}^{(r-1)})+(h_j^{(r+1)}-h_{j-1}^{(r+1)})+\\
+(\epsilon_j-\epsilon_{j+r-1})h_j^{(r)}+ \Upsilon_j^{(r)} b_j^{(r)} &=&0 \nonumber \\
(b_j^{(r-1)}-b_{j+1}^{(r-1)})+(b_j^{(r+1)}-b_{j-1}^{(r+1)})+\nonumber \\
+(\epsilon_j-\epsilon_{j+r-1})b_j^{(r)}- \Upsilon_j^{(r)} h_j^{(r)} &=&0 \nonumber,
\end{eqnarray}
where $\Upsilon_j^{(r)} \equiv 2\gamma+\Gamma \delta_{j,1}+\Gamma \delta_{j+r-1,L}$. There are $L^2$ equations for exactly as many unknown coefficients. Solving Eqs.(\ref{eq:eqs}) therefore gives exact expectation values of observables $H_j^{(r)}$ and $B_j^{(r)}$. 

The set of equations (\ref{eq:eqs0},\ref{eq:eqs}) can be compactly written in a matrix form by defining a hermitian correlation matrix $C$ of size $L \times L$ with matrix elements $C_{j,k} \equiv h^{(k-j+1)}_j+\ii\, b_j^{(k-j+1)}$ for $k>j$, $C_{j,j}\equiv h_j^{(1)}$, while $C_{j,k}=C_{k,j}^*$ for $j>k$. Let us define also $A \equiv \ii E-\ii J+\Gamma R$, where $E$ is a diagonal disorder matrix, $E_{j,j}=\epsilon_j$, while nonzero matrix elements of $J$ are $J_{i,i+1}=1, J_{i+1,i}=1$, while $R$ has only two nonzero elements, $R_{1,1}=R_{L,L}=1$. All $L^2$ equations (\ref{eq:eqs0},\ref{eq:eqs}) can now be written as a single matrix equation,
\begin{equation}
AC+CA^\dagger+2\gamma \tilde{C}=P,
\label{eq:C}
\end{equation}
where $P_{1,1}=-2\Gamma$, $P_{L,L}=2\Gamma$, while all other elements of $P$ are zero, and $\tilde{C}=C-{\rm diag}(C)$ is the correlation matrix without the diagonal. Physically, $P$ represents the driving, $E$ the disorder, $R$ the coupling to baths, $\tilde{C}$ term the dephasing while $J$ is due to XX Hamiltonian. Note that without dephasing, $\gamma=0$, matrix equation (\ref{eq:C}) would be of the Lyapunov type. Lyapunov equations appear in NESS solutions of quadratic systems, see e.g.~\cite{Lyapunov}.

Equation (\ref{eq:C}) is the starting point for our study. In some cases we are able to solve it analytically, in other we resort to numerically exact solution using standard linear algebra packages that enable solution for chain lengths $L$ of upto several thousand. 

Expectation values of all operators $H_j^{(r)}$ and $B_j^{(r)}$ are trivially proportional to $\mu$, see Eq.~(\ref{eq:ansatz}), and we therefore without loss of generality from now on set $\mu=1$. Coupling strength to reservoirs $\Gamma$ in general influences boundary resistance. Because we are interested in the regime of diffusive bulk transport, for which the boundary resistance does not matter in the thermodynamic limit, we in addition set $\Gamma=1$. For transport properties special attention shall be paid to the diagonal elements of the correlation matrix $C$, giving the magnetization profile, and to the imaginary part of the first near-diagonal, $b_j^{(2)}$, giving the current. Note that current is due to Eq.~(\ref{eq:eqs0}) independent of the site index; henceforth we shall frequently use a shorter notation $b \equiv b_j^{(2)}$, resulting in spin current expectation value $\ave{j}=4b$. Spin conductivity (depending on the context also called diffusion constant), being a proportionality constant between the gradient of a driving field, in our case magnetization, and a current, is therefore $\kappa=\lim_{L \to \infty}{\frac{L \ave{j_k}}{\ave{\sz_L-\sz_1}}}=\lim_{L \to \infty} 2L b$. Here we used that in the thermodynamic limit $\ave{\sz_L-\sz_1} \to 2$ (for $\mu=1$). We see that the conductivity is finite and nonzero, i.e., we have a diffusive transport, provided the current scales as $j \sim 1/L$.

Let us finally briefly discuss the energy range to which the studied NESS states correspond. We prefer to use energy instead of temperature because thermalization in integrable systems, of which the XX chain is an instance, can depend on the choice of Lindblad reservoirs~\cite{PRE:10}. There are two contributions to the energy (\ref{eq:H}), hopping and magnetization. As we shall see, the average magnetization profile is linear. Each individual term $\epsilon_i \sz_i$ will have fluctuations of order $\sim \sigma \mu$ centered around zero, meaning that their sum gives negligible contribution to the energy density in the thermodynamic limit. Similar conclusion holds for $\sx_j \sx_{j+1}+\sy_j \sy_{j+1}$. Namely, such term is essentially given by $\approx b^{(2)}_j (\epsilon_j-\epsilon_{j+1})/\gamma$ (see e.g. Eq.~\ref{eq:3}), again giving negligible contribution to the energy density in the thermodynamic limit. NESS solution (\ref{eq:ansatz}) is in the thermodynamic limit therefore always close to the identity matrix (i.e., an infinite temperature state). This means that we are studying transport properties at high energies.

\section{Fermionic picture}
It is helpful to translate quantities from spin language of Pauli matrices to the fermionic language of creation $c^\dagger_j$ and annihilation $c_j$ operators, satisfying anticommutators $\{c_j,c_k\}=0$, $\{c_j^\dagger,c_k^\dagger\}=0$, and $\{c_j,c_k^\dagger\}=\delta_{j,k}\mathbbm{1}$. Using Jordan-Wigner transformation~\cite{JW}, $c_j= -(\sigma_1^{\rm z}\cdots \sigma_{j-1}^{\rm z})\sigma_j^+$, or its inverse, $\sigma_j^{\rm x}= -(\sigma_1^{\rm z}\cdots \sigma_{j-1}^{\rm z})(c_j+c_j^\dagger)$, $\sigma_j^{\rm y}= -\ii (\sigma_1^{\rm z}\cdots \sigma_{j-1}^{\rm z})(c_j-c_j^\dagger)$, and $\sigma_j^{\rm z}= c_j c_j^\dagger-c_j^\dagger c_j=\mathbbm{1}-2n_j$, 
where we denote by $n_j=c_j^\dagger c_j$ a number operator at site $j$, the Hamiltonian (\ref{eq:H}) can be rewritten as a nearest-neighbor hopping, 
\begin{equation}
{\cal H}=\sum_j 2(c^\dagger_j c_{j+1}-c_j c_{j+1}^\dagger)+\sum_i \epsilon_i (1-2n_i).
\end{equation}
Random magnetic fields $\epsilon_i$ translate to a diagonal on-site disorder. Spin current $j_r$ translates to $j_r=4\ii (c^\dagger_r c_{r+1}+c_r c_{r+1}^\dagger)$, i.e., is proportional to the particle current. Bath Lindblad operators (\ref{eq:Lbath}) on the other hand inject or absorb a particle, thereby inducing nonzero particle current through the system. Operators $H_j^{(r)}$ and $B_j^{(r)}$ in the ansatz (\ref{eq:ansatz}) translate for $r \ge 2$ to
\begin{eqnarray}
H^{(r+1)}_j &=& 2(c_j^\dagger c_{j+r}-c_j c_{j+r}^\dagger) \nonumber \\
B^{(r+1)}_j &=& 2\ii (c_j^\dagger c_{j+r}+c_j c_{j+r}^\dagger).
\end{eqnarray}
They are two-point operators involving two fermions. Coefficients in the correlation matrix $C$ (\ref{eq:C}) therefore give expectation values of all two-point functions in the fermionic picture. In particular, density profile is $\ave{n_j}=(1+h_j^{(1)})/2$, while the particle (charge) current $j_k^{({\rm n})}$, defined as $j_k^{({\rm n})}=2\ii \, [n_k,c^\dagger_k c_{k+1}-c_k c_{k+1}^\dagger]=B_k^{(2)}$, has expectation value,
\begin{equation}
j\equiv \ave{j_k^{({\rm n})}}=2b.
\end{equation}
Conductivity in fermionic picture is therefore
\begin{equation}
\kappa=\lim_{L \to \infty}{\frac{L \ave{j_k^{({\rm n})}}}{\ave{n_L-n_1}}}=\lim_{L \to \infty}L 2 b,
\label{eq:kappa}
\end{equation}
and is the same as in spin language. In most of figures we shall either show the dependence of particle current $j_k^{({\rm n})}$, simply denoted by $j$, or of conductivity $\kappa$.

\section{Homogeneous disorder}
In this section we study the simplest case in which the disorder is present at each site and has the same variance. Each $\epsilon_j$ is an independent random variable with a uniform distribution $p(w)={\rm const.}$ in the interval $w \in [-\sqrt{3} \sigma,\sqrt{3}\sigma]$. We therefore have disorder averaged values $\ave{\epsilon_j}=0$ and $\ave{\epsilon_j^2}=\sigma^2$. Our results presented do not depend on details of the distribution $p(w)$, only on its width $\sigma$. Also, a homogeneous nonzero average $\ave{\epsilon_j}$ would not change any of the results presented, see discussion in Ref.~\cite{PRE11}.

In the next subsection we shall first study the case without dephasing for which one has Anderson localization (in the master equation setting). This regime is studied just to set the relevant length scale of localization. New results about the interplay of dephasing and disorder are then presented in the subsequent subsection~\ref{sec:B}.

\subsection{Localized phase, $\gamma=0$}
Without dephasing, $\gamma=0$, we have an ordinary tight-binding model with diagonal disorder in a non\-equilibrium setting. One expects that Anderson localization~\cite{Anderson} of the conservative model, for which in 1D all eigenstates of ${\cal H}$ are localized, will be reflected in an exponential decay of current with length $L$ (at constant driving). This has been numerically observed in Ref.~\cite{NJP10}; here we determine the decay rate, i.e., the localization length, because it is an important length scale relevant also for $\gamma \neq 0$. 
\begin{figure}[!h]
\centerline{\includegraphics[angle=0,scale=0.5]{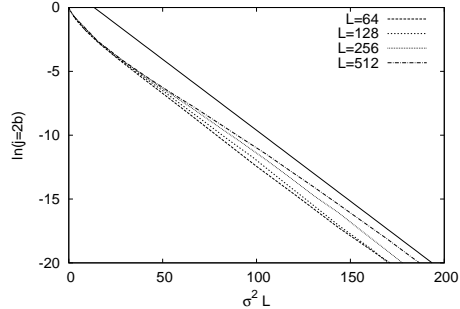}}
\caption{An ensemble averaged logarithm of the particle current for different chain lengths $L=64,128,256,512$ (bottom to top) and disorder strengths $\sigma$, all for $\gamma=0$. We have a universal exponential decay due to localization, $\ln{j} \sim (-L/l)$, with localization length $l \approx 9/\sigma^2$ (solid line above numerical curves). Non-asymptotic behavior for $\sigma^2 L \lesssim 10$ is due to localization length being larger than $L$.}
\label{fig:tokg0}
\end{figure}
In Fig.~\ref{fig:tokg0} we can see that the localization length determined from the dependence of current scales as $l \approx 9/\sigma^2$, consistent with previous similar observations in a nonequilibrium setting~\cite{Zilly:09}.
\begin{figure}
\centerline{
\includegraphics[angle=0,scale=0.34]{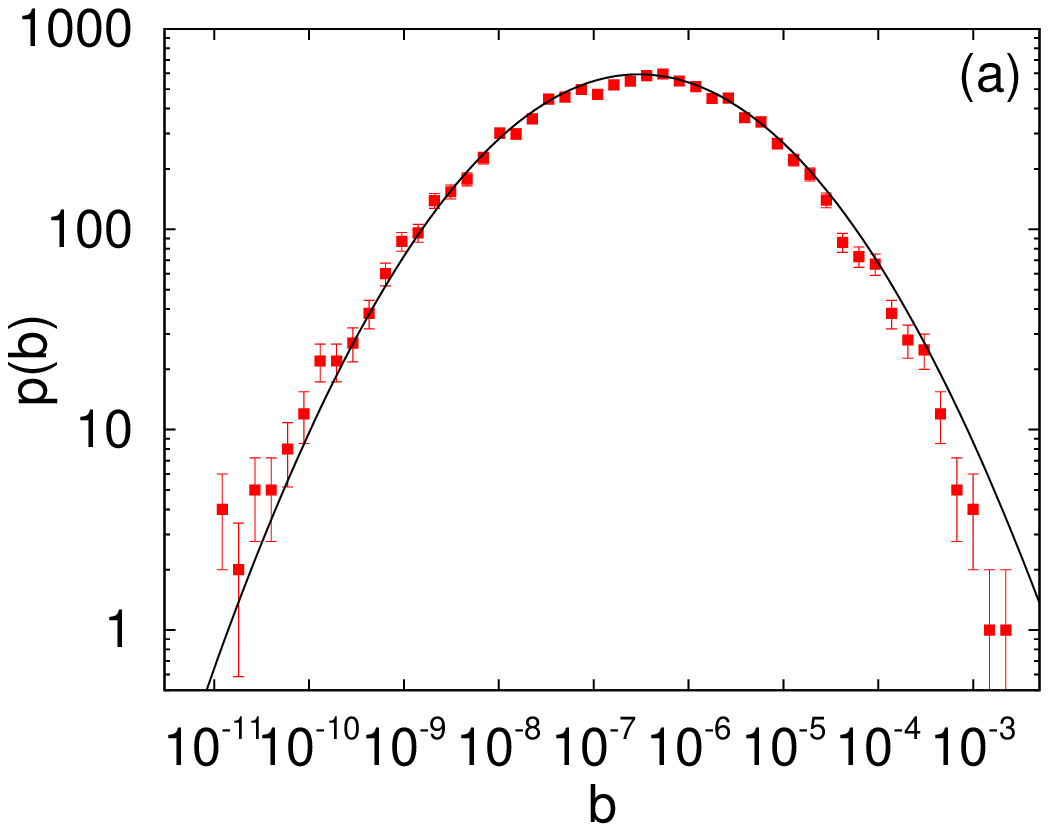}
\includegraphics[angle=0,scale=0.34]{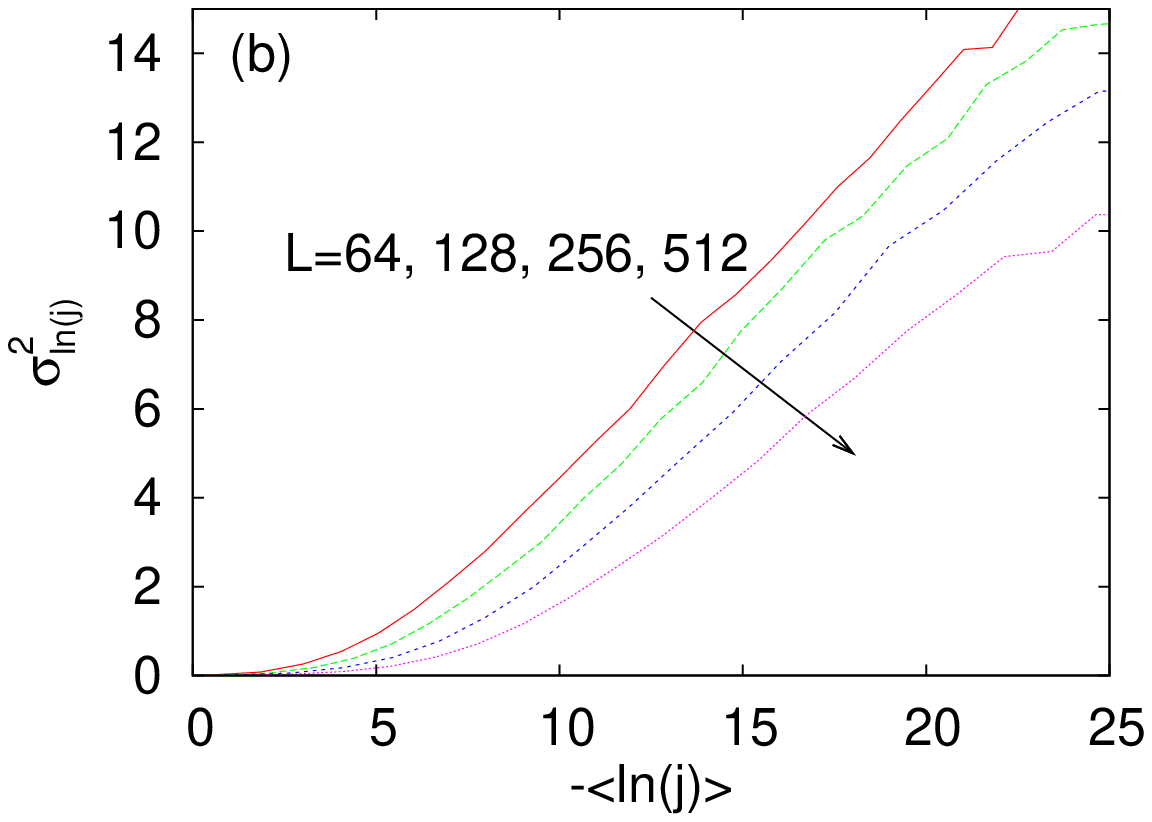}}
\caption{(a) Distribution of particle current (points) is close to log-normal (full curve). All is for $L=128$, $\sigma=1$. (b) Dependence of the variance of $\ln{j}$ on its average. Both plots are for zero dephasing, $\gamma=0$.}
\label{fig:stokg0}
\end{figure}
It is known from early studies of the Anderson model~\cite{Kramer:93} that the distribution of current is log-normal, i.e., distribution of its logarithm is normally distributed, away from band edges. For log-normal distribution the average current can be expressed as $\langle j \rangle=\exp{(m+s^2/2)}$, where $m=\langle \ln{j}\rangle, s=\sigma_{\ln{j}}$. The question of conductance distribution in the Anderson model at the band center or at the band edge however is a rather delicate one. Namely, an otherwise universal single-parameter scaling theory~\cite{Anderson:80}, saying essentially that the distribution function of conductance is a function of a single parameter (say of its first moment), is modified in the band center~\cite{Schomerus:03,Deych:03,Zilly:12}. We can see in Fig.~\ref{fig:stokg0} that in our nonequilibrium setting the distribution of current is almost log-normal. The single-parameter scaling relation between the first two moments of $\ln{j}$ on the other hand does not hold. Whether small discrepancies visible in the decay of current in Fig.~\ref{fig:tokg0} are due to the band-center anomaly is not clear.

\subsection{Diffusive regime for nonzero dephasing}
\label{sec:B}
We proceed to the main part of our work, that is to the case of nonzero dephasing $\gamma$. Relevant parameters are disorder strength $\sigma$, dephasing strength $\gamma$ and the chain length $L$. Fixing $\sigma$ and $\gamma>0$, we find that the current always scales diffusive as $j \sim 1/L$ for sufficiently long chains. That is, in the thermodynamic limit a nonzero dephasing always causes the system to become diffusive, regardless whether it was ballistic (for $\sigma=0$) or localized (for $\sigma \neq 0$) without dephasing. We note that interaction between particles still preserves diffusive nature of transport~\cite{NJP10}. We shall always discuss only properties in the thermodynamic limit, that is in the regime when the system is diffusive. For finite size effects that appear in the transition region from a localized/ballistic phase for small $L$'s to a diffusive phase see Appendix~\ref{app:finite}.

We first focus on the case of large disorder $\sigma$. For large $\sigma$ the size of $|C_{j,j+r}|$ exponentially decreases with the distance $r$ from the diagonal (we will see that this holds also for large $\gamma$). In the extreme case we can therefore approximate $C$ with a tridiagonal matrix: we assume that the only nonzero elements of $C$ are on the diagonal and in the 1st off-diagonal. Doing this approximation Eq.(\ref{eq:C}) can be solved exactly for any $L$, $\sigma$ and $\gamma$, see Appendix~\ref{app:3diag}. In the thermodynamic limit the expression for $b$ is
\begin{equation}
b=\frac{\gamma}{L}\frac{2}{2\gamma^2+\sigma^2},
\label{eq:b3}
\end{equation}
immediately giving conductivity (\ref{eq:kappa}),
\begin{equation}
\kappa = \frac{2}{\gamma+\frac{\sigma^2}{2\gamma}}.
\label{eq:kappa3}
\end{equation}
Expression of the same form has been obtained in Ref.~\cite{Hoyer:10} when studying time-evolution in a single-particle sector of a tight-binding model. Analogous formulas for the diffusion constant have been obtained even before in the context of spreading of particles in electric field~\cite{Bloch}, where the role of the disorder strength is played by the Bloch oscillation frequency.

\begin{figure}[!h]
\centerline{\includegraphics[angle=0,scale=0.5]{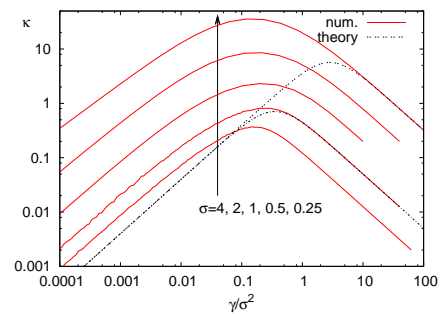}}
\caption{Conductivity at smaller disorder. Large-$\sigma$ theory (dotted curves), Eq.~(\ref{eq:kappa3}), still works for sufficiently large dephasing $\gamma \gtrsim \sigma$.}
\label{fig:small}
\end{figure}

\begin{figure}[!h]
\centerline{\includegraphics[angle=0,scale=0.5]{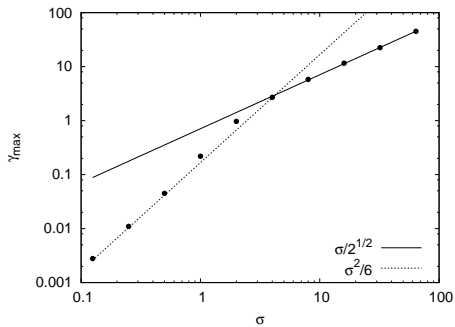}}
\caption{Location of the maximum $\gamma_{\rm max}$ in $\kappa(\gamma)$ as a function of disorder strength $\sigma$. Lines are asymptotic theoretical scaling, having two regimes, points are numerics.}
\label{fig:max}
\end{figure}

Transport properties in condensed-matter systems are many times discussed in terms of scattering lengths. Heuristically, one often assumes that different scattering processes are independent and that one can simply sum-up scattering rates, i.e., reciprocal scattering lengths. Interpreting analytical result for conductivity (\ref{eq:kappa3}) in such a way one can say that the total scattering length $l_{\rm tot}$ has an independent contribution from dephasing $l_{\rm deph}=1/\gamma$ and disorder $l_{\rm dis}=2\gamma/\sigma^2$, giving $1/l_{\rm tot}=1/l_{\rm deph}+1/l_{\rm dis}$ with $\kappa \propto l_{\rm tot}$. This of course holds only in the regime of validity of Eq.(\ref{eq:kappa3}), that is for large $\sigma$ or large $\gamma$.
\begin{figure}[!h]
\centerline{\includegraphics[angle=0,scale=0.5]{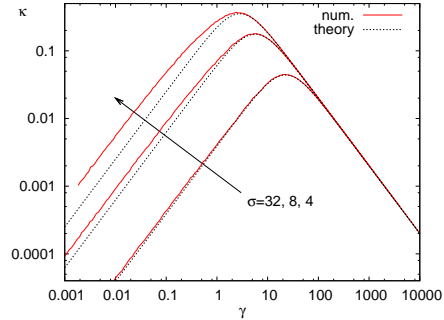}}
\caption{(Color online) Dependence of conductivity $\kappa$ on dephasing strength $\gamma$ for large disorder $\sigma$. Numerical results (full red curve) agree well with theoretical formula, Eq.(\ref{eq:kappa3}) (dotted curves).} 
\label{fig:n256}
\end{figure}
In Fig.~\ref{fig:n256} we can see that formula (\ref{eq:kappa3}) agrees well with numerical results for large $\sigma$. Two gross features are visible: (i) $\sim 1/\gamma$ dependence of $\kappa$ for large dephasing -- as the dephasing becomes large, scattering due to impurities can be neglected, and (ii) for small dephasing the transport is dominated by a break-down of localization caused by nonzero dephasing; correspondingly, $\kappa$ increases with the dephasing. A combined effect of the two contributions causes a maximum in conductivity at an intermediate dephasing strength. Similar behavior has been observed in other studies~\cite{Thouless:81,Zilly:09,Zilly:12,Stegmann:12} of dephasing effects on transport in a disordered tight-binding model as well as in transport properties of molecules or molecular aggregates~\cite{Segal:00,Nozaki:12,Plenio:08,Rebentrost:09,Kassal:12}. The location of the maximum in $\kappa$ scales according to (\ref{eq:kappa3}) as $\gamma_{\rm max}=\sigma/\sqrt{2}$, with the value of $\kappa$ at the maximum being $\kappa_{\rm max}=\sqrt{2}/\sigma=1/\gamma_{\rm max}$. Around $\sigma \approx 4$ and smaller, appreciable differences between numerical results and the theoretical Eq.~(\ref{eq:kappa3}) are visible. The reason for this failure at smaller $\sigma$ is that the approximation of the correlation matrix by a tridiagonal matrix becomes increasingly worse. For even smaller $\sigma$ it completely fails. Numerical results in this regime of $\sigma$ are shown in Fig.~\ref{fig:small}.
As opposed to large $\sigma$, the location of the maximum in the regime of small $\sigma$ scales as $\gamma_{\rm max} \sim \sigma^2$, Fig.~\ref{fig:small}, while the conductivity at the maximum is $\kappa_{\rm max} \sim 1/\sigma^2 \sim 1/\gamma_{\rm max}$. We see that the scaling of $\gamma_{\rm max}$ changes with $\sigma$. This can be nicely seen in Fig.~\ref{fig:max}. 
The crossover between the two asymptotic behaviors happens around $\sigma \approx 3$. We observe that this coincides with the disorder strength at which localization length for $\gamma=0$ becomes $l \approx 1$.

Observe in Fig.~\ref{fig:small} also that formula (\ref{eq:kappa3}) still holds for sufficiently large $\gamma$, i.e., for dephasing larger than about $\approx \sigma$. For $\gamma < \sigma$ and small $\sigma$, where (\ref{eq:kappa3}) does not work anymore, we were not able to obtain theoretical prediction for the current and $\kappa$. Including more that three diagonals in the calculation does not bring a significant improvement on Eq.(\ref{eq:kappa3}). It seems that the problem is in this regime strongly dominated by fluctuations and all elements of $C$ have to be taken into account. Perturbation theory in either $\gamma$ or $\sigma$ is also not successful, see discussion at the end of Appendix~\ref{app:finite}. In Fig.~\ref{fig:smallH} we show in more detail this small-$\sigma$ behavior.
\begin{figure}[!h]
\centerline{\includegraphics[angle=0,scale=0.5]{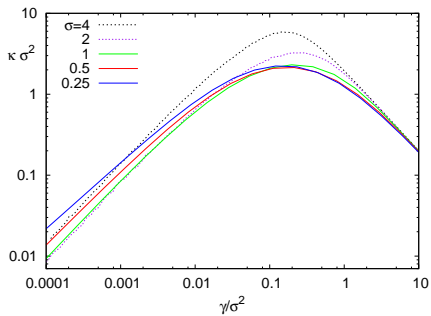}}
\caption{(Color online) Scaled conductivity $\kappa$ for small disorder $\sigma$.}
\label{fig:smallH}
\end{figure}
We can see that the dependence on $\gamma$ is rather complicated to the left of the maximum at $\gamma_{\rm max}$. Overall, $\kappa$ scales as $\sim 1/\sigma^2$ and is approximately a function of the scaled variable $\gamma/\sigma^2$. Scaling though is not perfect (see the region left of the maximum), perhaps pointing to a nontrivial (non-additive) interplay of dephasing and disorder. Whether this complicated behavior for small $\sigma$ is in any way related to anomalous properties of the Anderson model at the band center~\cite{diagE0a,diagE0b} (i.e., at zero energy) for weak uncorrelated disorder remains to be explored. Our Lindblad reservoirs namely induce a NESS with energy close to 0, i.e., in the band center.

\section{Diluted disorder}
In this section we discuss the case when disorder is absent at some sites. Such situation of a diluted disorder is relevant for real materials that can be prepared with different concentration of impurities. On theoretical side, it is known that in a Hamiltonian system non-homogeneous disorder, for instance a correlated one, can have a profound effect on the Anderson localization. In 1D correlations in disorder can namely cause a delocalization, see e.g.~\cite{1Ddeloc}. It is also worth mentioning that the band center can be special in terms of localization properties for weak (uncorrelated) diagonal~\cite{diagE0a,diagE0b} as well as for off-diagonal disorder~\cite{E0}.

We shall study two cases: (i) randomly placed diluted disorder in which a fraction $p$ of randomly chosen sites have disorder, and (ii) periodically placed disorder for which disorder is placed at every $\lambda$-th site. We will see that the transport properties of case (i) are very similar to the case of a homogeneous disorder with a renormalized disorder strength $\sqrt{p}\sigma$. Case (ii) though is qualitatively different. Disorder distribution function shall in all cases be the same uniform distribution as before.

\begin{figure}[!h]
\centerline{\includegraphics[angle=0,scale=0.5]{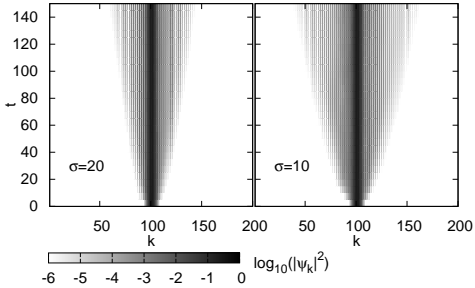}}
\caption{Wave-packet spreading in the case of periodic disorder ($\lambda=2$). Vertical axis is time, horizontal the spatial position within a system of length $L=200$. Due to a fraction of delocalized states the packet, initially localized around $k=100$, spreads with time. Two different disorder strengths shown, $\sigma=20$ and $\sigma=10$, result in different speeds of spreading. Note that the amplitude of the spreading part of a wavepacket is small (grayscale is logarithmic).}
\label{fig:speed}
\end{figure}

Before going to the properties of nonequilibrium states we briefly comment on the localization properties of a corresponding conservative system. Calculating eigenvalues and eigenstates of a tridiagonal matrix describing a tight-binding model with disorder at randomly placed fraction $p$ of sites, case (i), all eigenstates are still localized. Their localization length though differs. A fraction $p$ of eigenstates has localization length that is the same as if disorder would be at every site, while $(1-p)L$ eigenstates have larger localization length and are clustered around the band center at $E=0$. For periodic disorder, case (ii), delocalized states appear at the band center~\cite{Hilke97}. We found that all eigenstates with eigenenergies $|E| \lesssim \frac{1}{L \sigma}$ (for $\lambda=2$) are delocalized. Such delocalized eigenstates cause the spreading of initial localized packets as can be seen in Fig.~\ref{fig:speed}. We expect that this fraction $\sim 1/L$ of all eigenstates will strongly influence transport properties out of equilibrium if dephasing is zero or small.

\subsection{Randomly placed diluted disorder}
\begin{figure}[!h]
\centerline{\includegraphics[angle=0,scale=0.5]{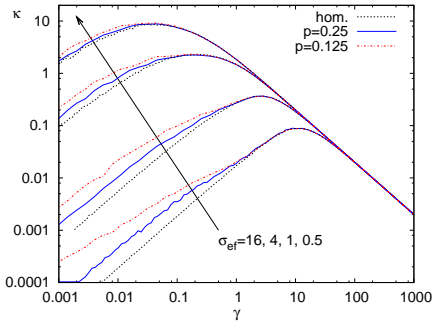}}
\caption{(Color online) Conductivity for a diluted disorder. Disorder is placed on random $pL$ sites. Black dotted curves are for a homogeneous disorder ($p=1$) with strength $\sigma_{\rm ef}$, blue and red are for a diluted case with $\sigma=\sigma_{\rm ef}/\sqrt{p}$.}
\label{fig:diluted}
\end{figure}

From the 3-diagonal derivation in Appendix~\ref{app:3diag}, holding for large $\gamma$, we see that  the fact that disorder is not present at each site will be reflected in an additional prefactor in front of $\sigma^2$. Instead of $L \sigma^2$ as for the homogeneous case we will have $pL\sigma^2$, resulting in the conductivity
\begin{equation}
\kappa = \frac{2}{\gamma+p\frac{\sigma^2}{2\gamma}}.
\label{eq:pkappa}
\end{equation}
The equation is the same as would be for the homogeneous case with an effective disorder strength $\sigma_{\rm ef}=\sqrt{p}\sigma$. This expression is expected to be valid for large $\gamma$. For smaller dephasing, even though we do not have a theoretical formula, we can still try to approximate the diluted situation with a homogeneous one at the effective disorder strength inferred from large $\gamma$ behavior (\ref{eq:pkappa}). In Fig.~\ref{fig:diluted} we compare results for a diluted disorder with the ones for a homogeneous case with a smaller effective disorder. The agreement between $\kappa$ for a randomly diluted disorder and a homogeneous one with a rescaled effective disorder is not perfect to the left of the maximum.

That for diluted disorder the disorder strength is just renormalized is not very surprising. Such result follows from the assumption that scattering events are independent and therefore scattering length scales as $\sim 1/p$. Scaling of such form has been observed experimentally in 1D spin chain materials described by the isotropic Heisenberg model in which scattering event are due to impurities~\cite{Hlubek:11}. We expect that findings of the present manuscript for the XX chain would not qualitatively change in the presence of interaction, i.e., for the Heisenberg model.

\subsection{Periodic disorder}
In this subsection we place disorder of strength $\sigma$ at every $\lambda$-th site, i.e., at sites $j=k\lambda-1,\quad k \in \mathbbm{N}$, while there is no disorder at all other sites.
\begin{figure}[!th]
\centerline{
\includegraphics[angle=0,scale=0.34]{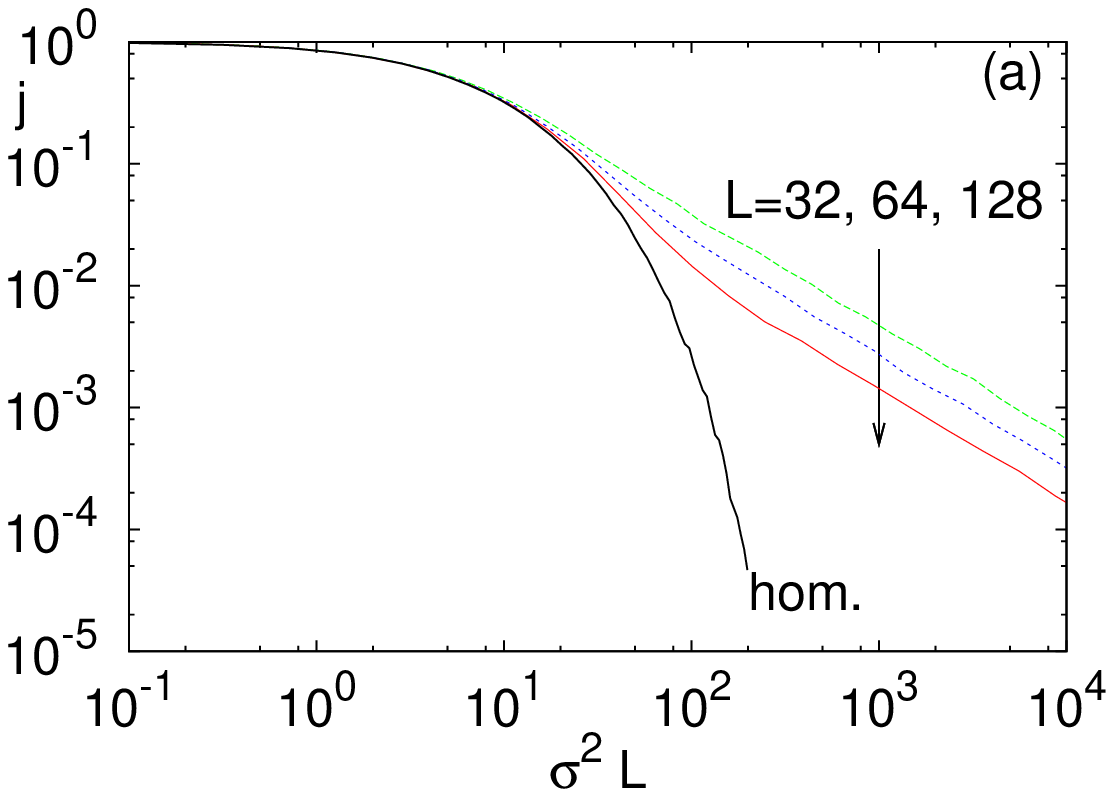}
\includegraphics[angle=0,scale=0.33]{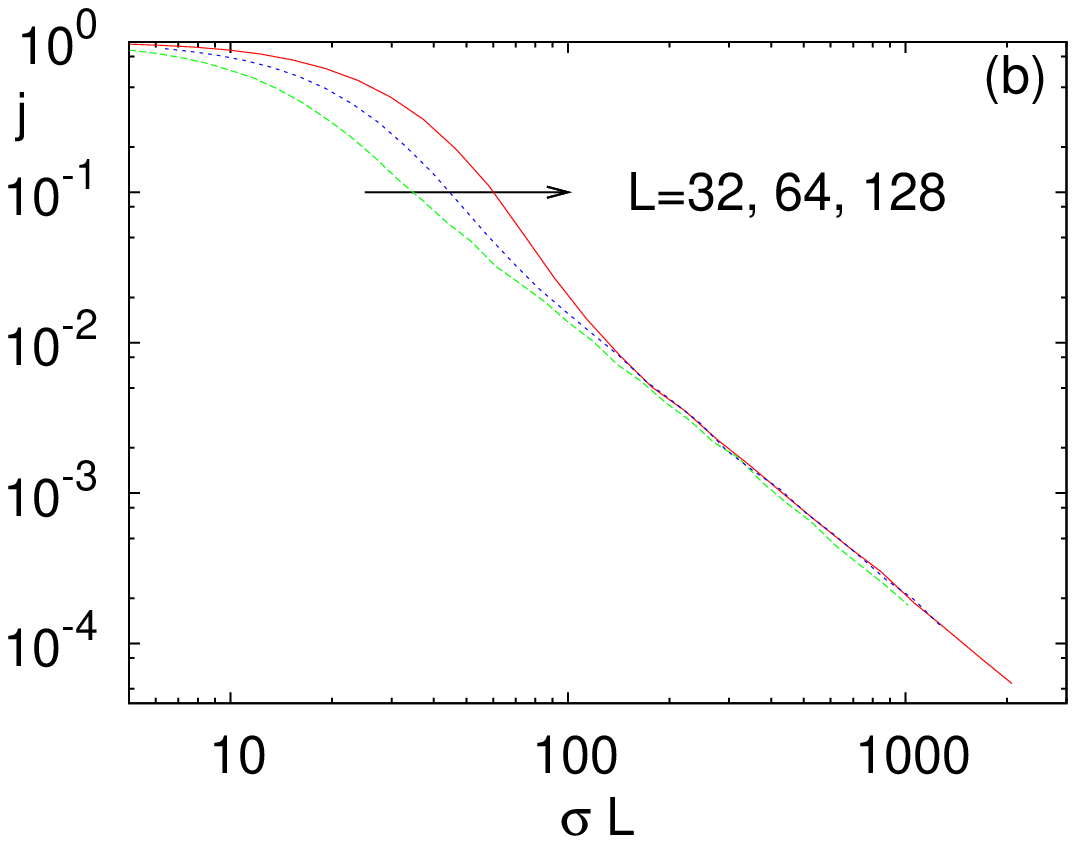}
}
\caption{(Color online) Periodic disorder with period $\lambda=2$ and $\gamma=0$. Scaling of the particle current $j$ with $\sigma$ for $L=32,64,128$. Difference between (a) and (b) is only in the scaling on the $x$-axis. (a) For $L \sigma^2 \lesssim 10$ (when localization length $l$ is larger than $L$) periodicity is irrelevant and the dependence is the same as for a homogeneous disorder with a rescaled disorder strength $\sigma/\sqrt{2}$ at every site (black solid curve ``hom.''; the same exponentially localized data as in Fig.~\ref{fig:tokg0}). Scaling variable is $\sigma^2 L$. (b) For larger disorder the scaling variable is $\sigma L$ and the decay is algebraic, $j \sim 1/(L \sigma)^\alpha$ with $\alpha \approx 2$.}
\label{fig:tokg0every2}
\end{figure}

First, we check conductivity for $\gamma=0$. Results of solving Eq.(\ref{eq:C}) and averaging over disorder are shown in Fig.~\ref{fig:tokg0every2}. For sufficiently small disorder $\sigma$, such that the localization length $l$ is larger than $L$, the dependence of current $j$ on $\sigma$ is the same as for a homogeneous disorder with a renormalized disorder strength of size $\sigma \sqrt{1/\lambda}$. One can see in Fig.~\ref{fig:tokg0every2}a that the differences between an exponentially localized homogeneous case and the periodic case begin to appear only for $\sigma^2 L \gtrsim 10$. For larger disorder, when periodicity becomes important, the nature of decay with $L$ changes (Fig.~\ref{fig:tokg0every2}b). Current begins to decays in an algebraic way as $j \sim 1/L^2$. Such decay can be traced back to properties of the Hamiltonian system, which has a fraction $1/L$ of delocalized eigenstates having eigenenergies of order $\sim 1/L$ located around the band center.

\begin{figure}[!h]
\centerline{\includegraphics[angle=0,scale=0.5]{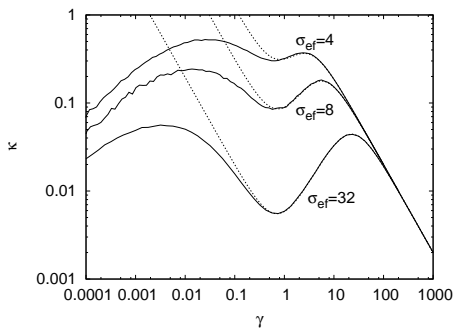}}
\caption{Conductivity for disorder on every 2nd site (full curves) and the theoretical Eq.~(\ref{eq:b5}), dotted curves.}
\label{fig:every2}
\end{figure}
We next move to nonzero dephasing for which the transport is always diffusive in the thermodynamic limit. System size at which this diffusive behavior is reached is larger than in the homogeneous case and depends on $\sigma$ and $\gamma$. Main difference with respect to the homogeneous or randomly diluted case is that the dependence of $\kappa(\gamma)$ has two peaks. An additional resonant peak appears at smaller $\gamma$ for sufficiently large disorder $\sigma$, see Fig.~\ref{fig:every2}. While the right peak is correctly described by Eq.~(\ref{eq:pkappa}) with $p=1/\lambda$, we have not been able to theoretically describe the left peak. Empirically we find that its position is 
\begin{equation}
\gamma_1 \approx 1/(3\lambda \sigma).
\label{eq:gamma1}
\end{equation}
Note that the position of the right peak is due to Eq.~(\ref{eq:pkappa}) 
\begin{equation}
\gamma_2 = \sigma/\sqrt{2\lambda}.
\label{eq:gamma2}
\end{equation}
The left peak is essentially a resonant phenomenon while the right one is due to competition between disorder and dephasing. Heuristically, $\gamma_1$ can be explained as the dephasing strength at which the dephasing time $1/\gamma$ is equal to the time a ballistic disturbance with speed $v$ needs to travel a distance $\lambda$. We have seen in Fig.~\ref{fig:speed} that for periodic disorder there are indeed delocalized eigenmodes in the absence of dephasing, with their speed being inversely proportional to $\sigma$, $v \sim 1/\sigma$. A resonant condition is therefore $1/\gamma \sim \lambda/v \sim \lambda \sigma$, resulting in Eq.~(\ref{eq:gamma1}).

\begin{figure}[!h]
\centerline{\includegraphics[angle=0,scale=0.5]{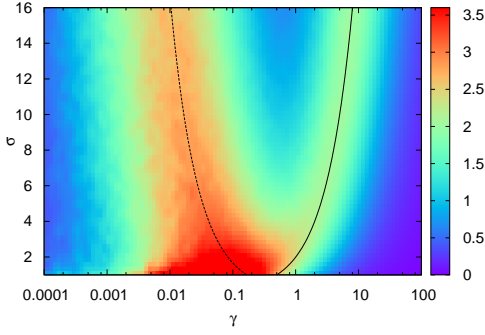}}
\caption{(Color online) Density plot of a scaled conductivity $\sigma\, \kappa$ as a function of $\gamma$ and $\sigma$ for disorder on every $\lambda=2$ site ($L=256$). We multiply conductivity by $\sigma$ in order to have better contrast for two peaks. Two black curves denote position of the maxima, left curve Eq.(\ref{eq:gamma1}), and right curve Eq.~(\ref{eq:gamma2}).}
\label{fig:n256scan}
\end{figure}

Even though we are not able to theoretically describe the left peak we can derive the location of the minimum between the two peaks. Assuming that nonzero elements of $C$ are only on the diagonal and two off-diagonals ($b^{(2)},h^{(2)}$ and $b^{(3)},h^{(3)}$), we can solve for $b$, see Appendix~\ref{app:5diag}. The result is in the thermodynamic limit
\begin{equation}
b = \frac{1}{L\gamma(1+\frac{\sigma_{\rm ef}^2}{2\gamma^2+1})},
\label{eq:b5}
\end{equation}
where $\sigma_{\rm ef}=\sigma/\sqrt{\lambda}$. Note that the difference from the 3-diagonal approximation (\ref{eq:b3}) is only in $1$ in the denominator. One can see in Fig.~\ref{fig:every2} that Eq.(\ref{eq:b5}) correctly describes the minimum. Location of the minimum is at $4\gamma_{\rm min}^2=\sigma_{\rm ef}^2-2-\sigma_{\rm ef}\sqrt{\sigma_{\rm ef}^2-8}$; this minimum is real for $\sigma_{\rm ef}>\sqrt{8}\approx 2.83$ while its location goes toward $\gamma_{\rm min} \asymp \sqrt{3}/2\approx 0.73$ for large $\sigma$. We therefore expect the left resonant maximum at $\gamma_1$ to appear only for $\sigma \gtrsim 3$ which is indeed confirmed by numerical calculation shown in Fig.~\ref{fig:n256scan}. From Eq.~(\ref{eq:b5}) we also see that the height at the minimum scales as $\asymp 1/\sigma^2$.
To also describe the maximum, and not just the minimum, we would have to solve for at least the diagonal and 4 off-diagonals in $C$, which proves to be too difficult (resonant maximum also appears only for chains with $L \ge 5$).
\begin{figure}[!h]
\centerline{\includegraphics[angle=0,scale=0.5]{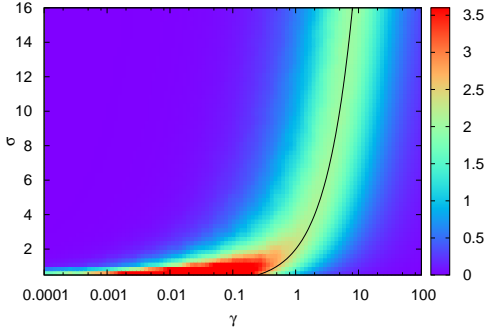}}
\caption{(Color online) Scaled conductivity $\sigma \kappa$ as a function of $\gamma$ and $\sigma$ for disorder at randomly chosen half of sites ($p=0.5$, $L=256$). Black curve is theoretical prediction for the location of the maximum at large $\sigma$, Eq.(\ref{eq:gamma2}) with $\lambda=2$.}
\label{fig:n256scanrd}
\end{figure}
In Fig.~\ref{fig:n256scan} we can see that locations of the two maxima is nicely predicted by Eqs.(\ref{eq:gamma1}) and (\ref{eq:gamma2}). Dependence of $\kappa$ for periodic disorder, having two maxima, can be compared to the one for randomly placed disorder (with the same average density), for which only the right maximum is present, see Fig.~\ref{fig:n256scanrd}.

\section{Conclusion}
We have studied conductivity in a tight-binding model with on-site disorder and experiencing dephasing. Transport is induced by a nonequilibrium driving described by the Lindblad master equation. Writing linear system of equations for all two-point fermionic expectation values in the steady state we study dependence of conductivity on dephasing and disorder strength. For large disorder or large dephasing we derive an analytic expression showing that the conductivity has a single maximum as a function of dephasing. We also discuss the case of diluted disorder. For randomly placed diluted disorder conductivity is approximately the same as for a homogeneous disorder with a rescaled disorder strength. For periodically placed disorder though a second resonant maximum in conductivity appears, exhibiting different scaling on parameters as the first maximum. We acknowledge support by the grant P1-0044.

\appendix

\section{3-diagonal approximation}
\label{app:3diag}
For large disorder $\sigma$ (and any $\gamma$) or for large $\gamma$ the size of correlations $|C_{j,j+r}|$ rapidly decreases away from the diagonal, that is, one has $|C_{j,j+r}| \sim \exp{(-r/\rho)}$ with small $\rho$. In such a case one can approximate $C$ by its diagonal $C_{j,j}$ and its two nearest diagonals, $C_{j,j+1}$, $C_{j+1,j}$. Setting all other matrix elements of $C$ to zero one can solve Eq.~(\ref{eq:C}) exactly as follows.

We use standard notation for the unknown terms in $C$, $h^{(1)}_j \equiv C_{j,j}$, $b \equiv \mathrm{Im}(C_{j,j+1})$ and $h^{(2)}_j \equiv \mathrm{Re}(C_{j,j+1})$. Two nontrivial equations (\ref{eq:C}) on the diagonal are
\begin{eqnarray}
2\Gamma h^{(1)}_1-2b&=&-2\Gamma \nonumber \\
2\Gamma h^{(1)}_L+2b&=&2\Gamma,
\label{eq:1}
\end{eqnarray}
while off-diagonal we have imaginary parts,
\begin{eqnarray}
b\, k+(h^{(1)}_1-h^{(1)}_2)+h^{(2)}_1(\epsilon_1-\epsilon_2) &=&0 \nonumber \\
2b \gamma+(h^{(1)}_2-h^{(1)}_3)+h^{(2)}_2(\epsilon_2-\epsilon_3) &=&0 \nonumber \\
\vdots \nonumber \\
2b \gamma+(h^{(1)}_j-h^{(1)}_{j+1})+h^{(2)}_j(\epsilon_j-\epsilon_{j+1}) &=&0 \nonumber \\
\label{eq:2}
\vdots \\
b\, k+(h^{(1)}_{L-1}-h^{(1)}_L)+h^{(2)}_{L-1}(\epsilon_{L-1}-\epsilon_L) &=&0,\nonumber
\end{eqnarray}
where $k\equiv \Gamma+2\gamma$, and real part,
\begin{eqnarray}
-k h^{(2)}_1+b(\epsilon_1-\epsilon_2) &=& 0 \nonumber \\
-2\gamma h^{(2)}_2+b(\epsilon_2-\epsilon_3) &=& 0 \nonumber \\
\vdots \nonumber \\
-2\gamma h^{(2)}_j+b(\epsilon_j-\epsilon_{j+1}) &=& 0 \nonumber \\
\vdots \nonumber \\
-k h^{(2)}_{L-1}+b(\epsilon_{L-1}-\epsilon_L) &=& 0.
\label{eq:3}
\end{eqnarray}
To get the coefficient of the current $b$ we sum all equations (\ref{eq:2}), and use the relation $h^{(1)}_1-h^{(1)}_L=2(b-\Gamma)/\Gamma$ obtained from (\ref{eq:1}). This gives
\begin{equation}
b(2k+(L-3)2\gamma)+\sum_{j=1}^{L-1} h^{(2)}_j(\epsilon_j-\epsilon_{j+1})+\frac{2}{\Gamma}(b-\Gamma)=0.
\label{eq:zvezab}
\end{equation}
It is worth pointing out that Eq.~(\ref{eq:zvezab}) holds true in general, not just in the case of a 3-diagonal approximation. Inserting into this relation expressions for $h^{(2)}_j$ obtained from Eqs.~(\ref{eq:3}), we get
\begin{equation}
b=\frac{2\gamma}{c+2\gamma^2 (L-1)+\sum_{j=2}^{L-2} (\epsilon_j-\epsilon_{j+1})^2/2},
\label{eq:3D-b}
\end{equation}
with the boundary term $c$ being independent of the length $L$, $c=2\gamma(\Gamma+\frac{1}{\Gamma})+\frac{\gamma}{\Gamma+2\gamma}\left[ (\epsilon_1-\epsilon_2)^2+(\epsilon_{L-1}-\epsilon_L)^2 \right]$. 

Let $\epsilon_j$ be distributed according to a distribution having a finite 2nd moment, $\ave{\epsilon_j^2}=\sigma^2$. Then $\epsilon_j-\epsilon_{j+1}$ has 2nd moment equal to $2\sigma^2$. According to the central limit theorem a sum of terms $(\epsilon_j-\epsilon_{j+1})^2$ will converge to a Gaussian distributed random variable with a nonzero average $2(L-3)\sigma^2$ and a variance $\propto L$. Expression $\sum_{j=2}^{L-2} (\epsilon_j-\epsilon_{j+1})^2/2$ therefore becomes increasingly sharply peaked about its average, with relative fluctuations being of order $\sim 1/\sqrt{L}$. Therefore, in the limit $L \to \infty$ when we can neglect boundary terms, we can write
\begin{equation}
b=\frac{2\gamma}{L(2\gamma^2+\sigma^2)}.
\end{equation}

\section{5-diagonal approximation for periodic disorder}
\label{app:5diag}
Let us have disorder only on odd sites, $\epsilon_{2j}\equiv 0$, and assume that the correlation matrix is 5-diagonal, $C_{j,j+r}\equiv 0$ for $r>2$. The goal would be to go beyond the 3-diagonal approximation in order to describe the 2nd resonance peak visible for instance in Fig.~\ref{fig:every2}. Although we will fail at that -- one would need to solve at least a 11-diagonal approximation to get the 2nd peak (which we are though not able to do analytically) -- we will nevertheless correctly describe the minimum between the two peaks giving us at least some insight.

We will neglect boundary effects and solve equations in the leading order in $\gamma/\sigma^2$. This is justified because the 2nd resonance maximum appears only for large $\sigma$. First, we note that because $\epsilon_{2j}=0$ one has $h^{(3)}_{2j}=0$. Then we also observe that $b^{(3)}_{2j+1} \sim \gamma/\sigma^2$ and can be neglected. We therefore start with the fact that $h^{(3)}$ are zero on even sites while $b^{(3)}$ are (approximately) zero on odd sites. On even sites we have $b_{2j}^{(3)}=-(h^{(2)}_{2j}-h^{(2j)}_{2j+1})/(2\gamma)$. Writing now equations for the real part of the near-diagonal at two consecutive sites in the bulk, we have
\begin{eqnarray}
-2\gamma h^{(2)}_{2j}-\frac{h^{(2)}_{2j}-h^{(2)}_{2j+1}}{2\gamma}-\epsilon_{2j+1} b &=& 0 \nonumber \\
-2\gamma h^{(2)}_{2j+1}+\frac{h^{(2)}_{2j}-h^{(2)}_{2j+1}}{2\gamma}+\epsilon_{2j+1} b &=& 0 \nonumber.
\end{eqnarray}
From these two equations we see that only two nearest-neighbor $h^{(2)}$ are coupled. Instead of $L$ coupled equations we have a set of uncoupled $2\times 2$ equations. Solving them we get
\begin{equation}
-h_{2j+1}^{(2)}=h_{2j}^{(2)}=-\frac{2\gamma}{(2\gamma)^2+2}\epsilon_{2j+1} b.
\end{equation}
Inserting these into Eq.(\ref{eq:zvezab}) we get in the thermodynamic limit $L \to \infty$,
\begin{equation}
b=\frac{2\gamma}{L(2\gamma^2+\sigma^2\frac{1}{1+1/(2\gamma^2)})}.
\label{eq:every2}
\end{equation}
This expression correctly describes the minimum between the two peaks in Fig.~\ref{fig:every2}, although it fails to reproduce the left maximum.

\section{Finite-size effects}
\label{app:finite}
Limits $L \to \infty$ and $\gamma \to 0$ do not commute. Depending on the order of the two we get either a diffusive behavior $j \sim 1/L$ or an insulating $j \sim a^{-L}$ due to localization. We are interested in the diffusive behavior obtained in the correct thermodynamic limit of letting $L \to \infty$ first. In order to observe diffusive behavior one therefore has to take sufficiently large $L$ for each particular $\gamma$.

\begin{figure}[!h]
\centerline{\includegraphics[angle=0,scale=0.5]{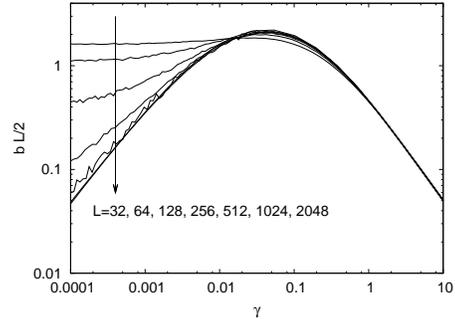}}
\caption{Scaled particle current for $\sigma=0.5$ and different sizes $L$. Convergence to diffusive scaling, seen as an overlap of curves, is for smaller $\gamma$ reached only for large sizes (data for $L=1024$ and $2048$ overlap and are indistinguishable on the scale of the plot).} 
\label{fig:znh05}
\end{figure}
\begin{figure}[!h]
\centerline{\includegraphics[angle=0,scale=0.5]{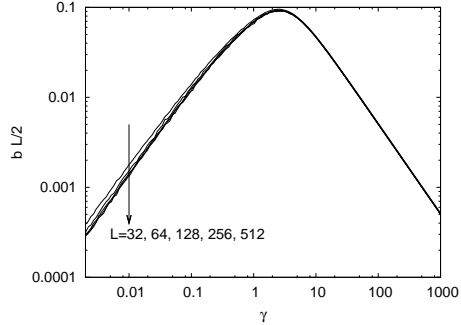}}
\caption{Same as Fig.~\ref{fig:znh05}, but for $\sigma=4$. Data for $L=128, 256, 512$ overlap.}
\label{fig:znh4}
\end{figure}
For instance, in Fig.~\ref{fig:znh05} we can see that at $\gamma=10^{-4}$ one needs $L=1024$ or larger, while at $\gamma=10$ chain length $L=32$ is enough to observe diffusive behavior. The system size $L$ at which diffusive behavior is reached is smaller for larger disorder strength $\sigma$, as can be seen by comparing Figs.~\ref{fig:znh4} and~\ref{fig:znh05}.

We also note that the perturbation theory in $\gamma$ around $\gamma=0$ has a convergence radius that shrinks to zero with the system size $L$. The failure of a naive perturbation theory to predict diffusive behavior at small nonzero dephasing is not very surprising because the nature of transport changes abruptly at $\gamma=0$. Similar are problems trying to apply perturbation theory in $\sigma$ in order to obtain behavior at small $\sigma$ and $\gamma$ (for instance, the one in Fig.~\ref{fig:smallH}).


\begin{thebibliography}{1}
\bibitem{foot1} Note that having a nonzero average magnetization in the driving, i.e. a nonzero $\bar{\mu}$ in Eq.(7-8) of Ref.~\cite{PRE11}, only shift the overal magnetization profile while it has no effect on the remaining $L(L-1)$ two-point expectations studied in the present work.
\end{thebibliography}
\end{document}